\documentclass[manuscript, screen = False, review = False, anonymous = False]{acmart}
\AtBeginDocument{%
  \providecommand\BibTeX{{%
    \normalfont B\kern-0.5em{\scshape i\kern-0.25em b}\kern-0.8em\TeX}}}

\usepackage{subcaption}
\usepackage{float}
\usepackage{natbib}
\usepackage{graphicx}
\usepackage{booktabs}
\usepackage{rotating}
\usepackage{lscape}
\usepackage{xspace}
\def\citepos#1{{\hypersetup{citecolor=black}\citeauthor{#1}}'s \citep{#1}}

\let\oldciteauthor=\citeauthor
\def\citeauthor#1{{\hypersetup{citecolor=black}\oldciteauthor{#1}}}

\setcopyright{acmcopyright}
\copyrightyear{xxxx}
\acmYear{xxxx}
\acmDOI{XXXXXXX.XXXXXXX}

\acmConference[Conference short name]{Conference}{Dates}{Location}
\acmBooktitle{acm booktitle} 
\acmPrice{price}
\acmISBN{978-1-4503-XXXX-X/18/06}

\newcommand{\gh}{GitHub\xspace}

\usepackage{xcolor}
\definecolor{darkgreen}{rgb}{0.05,0.5,0.05}

\usepackage{verbatim}

\begin{document}

\title[``It would work for me too'': 
How Online Communities Shape Software Developers' \\Trust in AI-Powered Code Generation Tools]{``It would work for me too'': 
How Online Communities Shape Software Developers' Trust in AI-Powered Code Generation Tools}

\author{Ruijia Cheng}
\authornote{This work is done during the author's internship at Microsoft Research.}
\email{rcheng6@uw.edu}
\affiliation{%
  \institution{University of Washington}
  \city{Seattle}
  \state{Washington}
  \country{USA}
  \postcode{98195}
}

\author{Ruotong Wang}
\authornotemark[1]
\email{ruotongw@cs.washington.edu}
\affiliation{%
  \institution{University of Washington}
  \city{Seattle}
  \state{Washington}
  \country{USA}
  \postcode{98195}
}

\author{Thomas Zimmermann}
\email{tzimmer@microsoft.com}
\affiliation{%
  \institution{Microsoft Research}
  \city{Redmond}
  \state{Washington}
  \country{USA}
  \postcode{98052}
}

\author{Denae Ford}
\email{denae@microsoft.com}
\affiliation{%
  \institution{Microsoft Research}
  \city{Redmond}
  \state{Washington}
  \country{USA}
  \postcode{98052}
}

\renewcommand{\shortauthors}{Cheng, et al.}

\begin{abstract}
While revolutionary AI-powered code generation tools have been rising rapidly, we know little about how and how to help software developers form appropriate trust in those AI tools. Through a two-phase formative study, we investigate how online communities shape developers' trust in AI tools and how we can leverage community features to facilitate appropriate user trust. Through interviewing 17 developers, we find that developers collectively make sense of AI tools using the experiences shared by community members and leverage community signals to evaluate AI suggestions. We then surface design opportunities and conduct 11 design probe sessions to explore the design space of using community features to support user trust in AI code generation systems. We synthesize our findings and extend an existing model of user trust in AI technologies with sociotechnical factors. We map out the design considerations for integrating user community into the AI code generation experience.   

\end{abstract}



\maketitle

\section{Introduction}

Recent years have witnessed a rapid development of generative AI models, with applications in areas like creative writing, digital arts, and text summarizations. Perhaps one of the most exciting areas of the application of generative AI models is software development with automatic code generation. Several AI-powered code generation tools have emerged recently---\gh Copilot\footnote{\url{https://github.com/features/copilot}}, Tabnine\footnote{\url{https://www.tabnine.com/}}, Kite\footnote{\url{https://www.kite.com/}}, Amazon Code Whisperer\footnote{\url{https://aws.amazon.com/codewhisperer/}}, ChatGPT\footnote{\url{https://openai.com/blog/chatgpt/}}---to name a few. These tools have been considered revolutionary for the workflow of software developers that is long considered complex and highly demanding \cite{sarkarWhatItProgram2022,ernst2022ieeesoftware}, 
as they are designed to augment developers' productivity by auto-completing lines of code in real time, and/or generating code based on prompts \cite{zieglerProductivityAssessmentNeural2022,barkeGroundedCopilotHow2022}. 
Despite being recently invented, AI-powered code generation tools have already gained a huge amount of attention and popularity. For the example of \gh Copilot, more than 1.2 million developers participated in its technical preview, and the user population is still fast growing since Copilot became generally available on June 21st, 2022 \cite{Dohmke2022}. 

Along with the vast popularity comes the concern on the developer's trust in the AI. Many strive to help developers form and maintain an \emph{appropriate} level of trust in AI code generation tools---
an insufficient level of trust can prevent developers from using AI to increase productivity \cite{murphyhill2021tse,storey2021tse}, while too much trust can lead to overlooking risks and safety vulnerabilities to software used by millions of users \cite{lipner2004,hasselbring2006, Zakir2014}. Since user trust in AI is shaped by how an user interact with an AI according to \citepos{liaoDesigningResponsibleTrust2022} \emph{MATCH model} (explained in §\ref{sec:background_trust}), and user trust in technology more broadly has long been considered as a construct situated in a social context \cite{campos-castilloSituatedTrustPhysician2019, childTrustFundamentalBond2001, harawaySituatedKnowledgesScience1988, zhangShiftingTrustExamining2022}, we believe that appropriate user trust in AI is \emph{sociotechnical}---a product of interaction \emph{between} developers \emph{about} their interaction with AI. 
However, we know little about how we can help developers build appropriate trust in AI in sociotechnical systems. 

Fortunately, online developer communities provide us with a promising angle to study how developers build (or face challenges to build) appropriate trust in AI-powered code generation tools. 
We have known from an abundance of HCI literature that developers thrive in online communities to ask and answer questions \cite{treude2011icse,barua2014developers}, learn to adopt new tools and skills, share projects and exchange feedback \cite{Chattopadhyay2021CSCW, Zagalsky2016msr, cheng2020building}, and seek technical and social support~\cite{ford2018dont}. 
Following the emergence of AI-powered code generation tools, tens of thousands of developers have been organically forming end-user online communities on platforms such as \gh, Stack Overflow, Reddit, and Twitter to share and discuss their interaction with AI---some of these discussions are productive, while others are not. As we have the rare opportunity to witness the onset of online developer communities for new AI technology, we study the role of online communities in facilitating developers to build trust with AI and the challenges that developers face in this process. We hope to learn from these communities to design for developers to build appropriate trust in AI code generation systems, as well as future AI end-user communities for developers.


We present a two-phase formative study to understand how developers build appropriate trust in AI-powered code generation tools in online communities and explore the design space. In Study 1, we conducted semi-structured interviews with 17 developers on their experience participating in online communities about AI code generation tools (§\ref{sec:study1}). We identified two types of information that developers can gain from online communities to build appropriate trust in AI: they can engage with \textit{community-curated experiences} that allow them to understand the capacities of AI, as well as leverage \textit{community-generated evaluation signals} to determine how to trust specific code suggestions. 
In Study 2, building on the developers' needs and challenges that we discovered from the interviews, we mapped out the design opportunities of introducing a user community to AI code generation systems. Specifically, we created visual stimuli that integrate \textit{community evaluation signals} and \textit{experience sharing spaces} into the GitHub Copilot experience, and presented the visual stimuli to 11 developers in a design probe study to collect further conceptual-level insights and ideas for design (§\ref{sec:study2}). To synthesize the findings of our formative studies, we extended \citepos{liaoDesigningResponsibleTrust2022} \emph{MATCH model} and discussed the two pathways that socio-technical systems can support users to build appropriate trust in AI---collective sensemaking and community heuristics (§\ref{sec:discussion_theory}). We also summarized several design recommendations for integrating a user community into AI systems to help build appropriate trust in AI (§\ref{sec:discussion_design}). 
 
We make the following contributions to the HCI community: 
(1) We contribute empirical insights that explain how online communities help developers build appropriate trust with AI code generation tools. (2) We extend the previous \emph{MATCH model} of user trust in AI technologies by adding the role of socio-technical systems in trust building. (3) Finally, we unveil a series of important and user-evaluated design concepts coupled with visual examples for the future development of AI code generation tools.

\section{Related Works} 
\subsection{Building appropriate trust in AI: the \emph{MATCH model}}
\label{sec:background_trust}
Trust has been highlighted as one of the most important factors in multiple design guidelines for human-AI collaboration ~\cite{amershiGuidelinesHumanAIInteraction2019, eu2019building}. Trust in AI is defined as the user's attitude or judgement about how the AI system can help when the user is in a situation of uncertainty or vulnerability~\cite{liaoDesigningResponsibleTrust2022, leeTrustAutomationDesigning2004, vereschakHowEvaluateTrust2021}, and is therefore particularly important when users engage in high-stake activities where the consequence can have severe impact ~\cite{jacoviFormalizingTrustArtificial2021}. For example, in software development, the trust of the user in AI support is important both for tool adoption ~\cite{shundan2014cscw,witschey2015} and for product safety \cite{lipner2004,hasselbring2006}. A misalignment between the trust of users and AI's ability could lead to user overreliance and misuse of AI systems, resulting in unwanted consequences ~\cite{passiOverrelianceAILiterature, dzindoletRoleTrustAutomation2003, leeTrustAutomationDesigning2004}. Therefore, in addition to making technologies more trustworthy, it is also important to support users in deciding when and how much to trust an AI, so that they can use their expertise when needed to complement the capabilities of AI systems~\cite{jacoviFormalizingTrustArtificial2021}.

\begin{figure*}[t]
  \centering
  \includegraphics[width=0.8\linewidth]{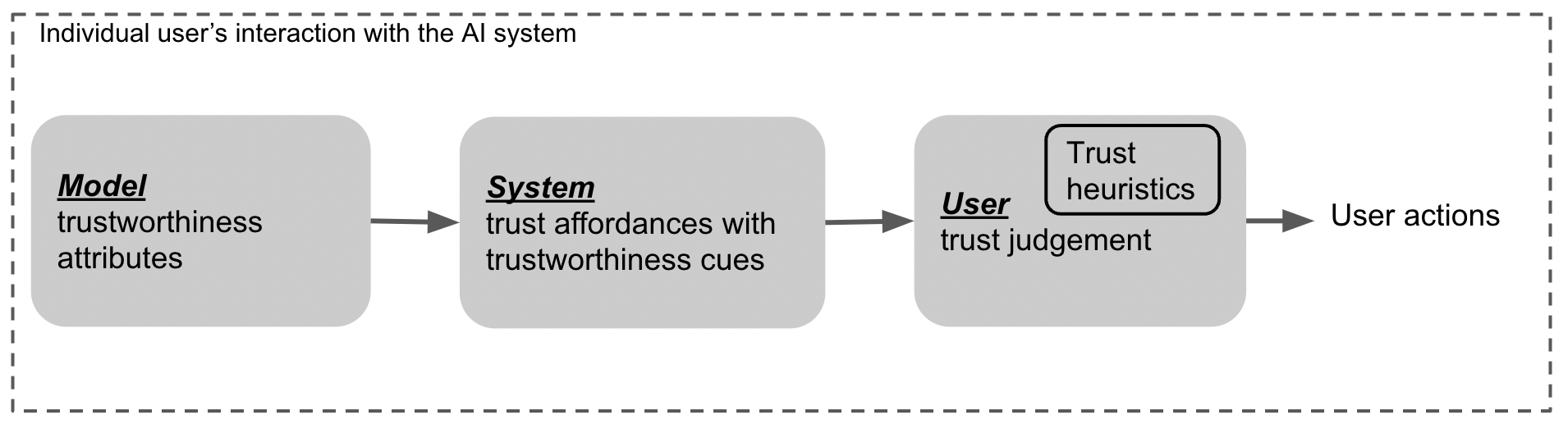}
  \caption{The \emph{MATCH model} proposed by Liao and Sundar~\cite{liaoDesigningResponsibleTrust2022}, which describes how trustworthiness is communicated in AI systems and how users make trust judgements.}
  \Description{A flowchart diagram of the MATCH model. Leftmost box with text trustworthiness attributes at the model level. Middle box with text System level trust affordances with trustworthiness cues. Rightmost box with text User level trust judgement. The end of the flow chat is text User actions.}
  \label{fig:match_model}
\end{figure*}
For this reason, the HCI community has been calling builders of AI systems to design for \emph{appropriate trust} that is calibrated to different use cases and scenarios~\cite{liaoDesigningResponsibleTrust2022, jacoviFormalizingTrustArtificial2021}. An important premise to support appropriate trust is understanding how trust is developed or impacted when a user is interacting with an AI system. Recently, \citet{liaoDesigningResponsibleTrust2022} presents the \emph{MATCH model} (as we visualize in Figure \ref{fig:match_model}) derived from a comprehensive review of the literature that explains the factors that affect how users build trust with an AI system ~\cite{liaoDesigningResponsibleTrust2022}. In this paper, we use the \emph{MATCH model} as the framework and source of vocabulary for us to describe how users build trust in AI technologies. We explain the \emph{MATCH model} as follows. 

An AI model is typically built with certain \textit{trustworthy attributes} (leftmost box in the figure), such as its ability, benevolence, and integrity, that can directly affect users' trust~\cite{yuTrustMyMachine2019, mayerIntegrativeModelOrganizational1995a, drozdalTrustAutoMLExploring2020, yinUnderstandingEffectAccuracy2019}. For example, in the case of a code generation AI, the trustworthy attributes can be its model attributes and how it was trained. These capabilities are communicated to users via \textit{trust affordances} with trustworthiness cues presented in the interface design (middle box). In the case of a code generation AI, the trust affordances can be how the AI-generated code is presented to users. Users then form \textit{trust judgements} (rightmost box) that result in their actions towards the AI system by interpreting those trustworthiness cues using their own \textit{trust heuristics}, such as their prior knowledge in the domain, perception of the reputation of the AI, emotion elicited by the system design, and so on. In the case of an AI system that generates code, this can happen when a developer uses their knowledge to decide whether to take the AI-generated code suggestion. 

The \emph{MATCH model} therefore has pointed out two potential pathways to support users to build appropriate trust with AI systems: communicating richer trustworthiness cues, and guiding users with trust heuristics. HCI researchers have begun to explore system designs to support these two pathways. For instance, ~\citet{drozdalTrustAutoMLExploring2020} found that transparency features in an AutoML tool can help users understand the system and build trust with the system. ~\citet{zhangEffectConfidenceExplanation2020a} found that an implementation of a confidence score can support users in calibrating trust. 

An important limitation of the current \emph{MATCH model} and relevant design is that it focuses on individual users' interaction with an AI system, while not explicitly considering the broader sociotechnical ecosystem surrounding the AI. As ~\citet{liaoDesigningResponsibleTrust2022} point out, extrinsic cues such as other users' reaction with the AI system can potentially affect the current users' trust judgement. \citet{ehsanExpandingExplainabilitySocial2021} discovered that users' perception and needs in AI are situated in a social, collaborative setting. Beyond the context of human-AI collaboration, a recent study has shown that trust is situated and shaped by users' information-seeking and assessment practices on online social platforms~\cite{zhangShiftingTrustExamining2022}. We explore the opportunities of leveraging online communities to help users build appropriate trust with AI-powered code generation tools. In other words, what people see about AI code generation tools in online communities and how they engage with it may affect their trust and online communities have the potential to support users make informed trust judgements.

\subsection{Developers' participation in online communities}
Online communities have served as a sociotechnical ecosystem for users to connect and engage with each other through a range of interactions~\cite{boyd2007social}.  As online communities evolve over time, there is a select set of values that users find to be core to their success. One of those core values is the opportunity to engage with the community. The ability to share experiences through posts, respond to the experiences of others through comments, and/or have the flexibility to engage at low-risk levels such as voting have become a common standard of what users expect~\cite{chen2018user}. This ability to engage with other members or even actively observe has solidified these virtual spaces as an integral setting for situated learning ~\cite{lave1991situated}. Another essential core value of online communities is the expectation of a variety of experiences being shared. For instance, as of 2018, Reddit is home to more than 138,000 active communities alone\footnote{\url{https://www.chicagotribune.com/business/ct-biz-reddit-chicago-office-20180418-story.html}}---topics ranging from understanding privacy and security of devices~\cite{krsek2022selfpersuade} to debating the impact of deep fakes~\cite{gamage2022deepfakes}. Gathering a peer perspective and also being able to inquire about new technologies with those who have a common interest but have never met before work is a rare opportunity that online communities provide.
Arguably, one of the most critical components of online communities is trustworthiness in the quality of content on the platform~\cite{weld2022makes}. This is especially critical when someone is sharing technical content on complex topics, such as scientific explanations ~\cite{jones2019challenges}. In these technical settings, trusting the quality of the source of the information and having confidence in the expertise of the members set the tone for how others decide to participate.

Online developer communities appear to have similar core values as well as caution. In communities such as Stack Overflow and \gh, builders of software tools convene to discuss new programming paradigms~\cite{mamykina2011design}, engage in debate on ethical considerations of new tools~\cite{widder2022limits}, and more often than not lurk until they have something worth adding to the conversation~\cite{steinmacher2015social}. More broadly, developers engage with other community members 
 to seek answers, feedback, and mentorship~\cite{treude2011icse,barua2014developers, ford2018dont}, as well as to discover new tools, resources, and inspirations~\cite{Chattopadhyay2021CSCW, Zagalsky2016msr, cheng2020building}. 
Throughout all these forms of engagement, community members are still learning by observing experts ~\cite{ford2018dont} and often trying to connect experiences shared back to their own~\cite{faas2018watch}. Developers have relied on these communities so much that researchers and toolsmiths have found it worthwhile to embed them into the development process. For instance, \citet{bacchelli2012harnessing} built a tool named Seahawk that experimented with accessing Stack Overflow snippets in the IDE (Integrated Development Environment) to reduce the friction between developers finding the resources they need, gathering, and contributing back to the developer community---all while staying productively on task.

Inspired by the affordances that communities can provide to software developers, we see the potential of online communities offering a promising setting for developers to learn and build trust with AI code generation tools. As applications of generative AI have been exploding in recent years, we have the opportunity to observe developers forming and leveraging communities to make sense of those AI technologies. We thus unfold this investigation to understand the impact of online communities on developers' trust in AI-powered code generation tools and what we can learn from these communities.

\subsection{AI-powered code generation tools}

\begin{figure*}[t]
  \centering
  \includegraphics[width=0.8\linewidth]{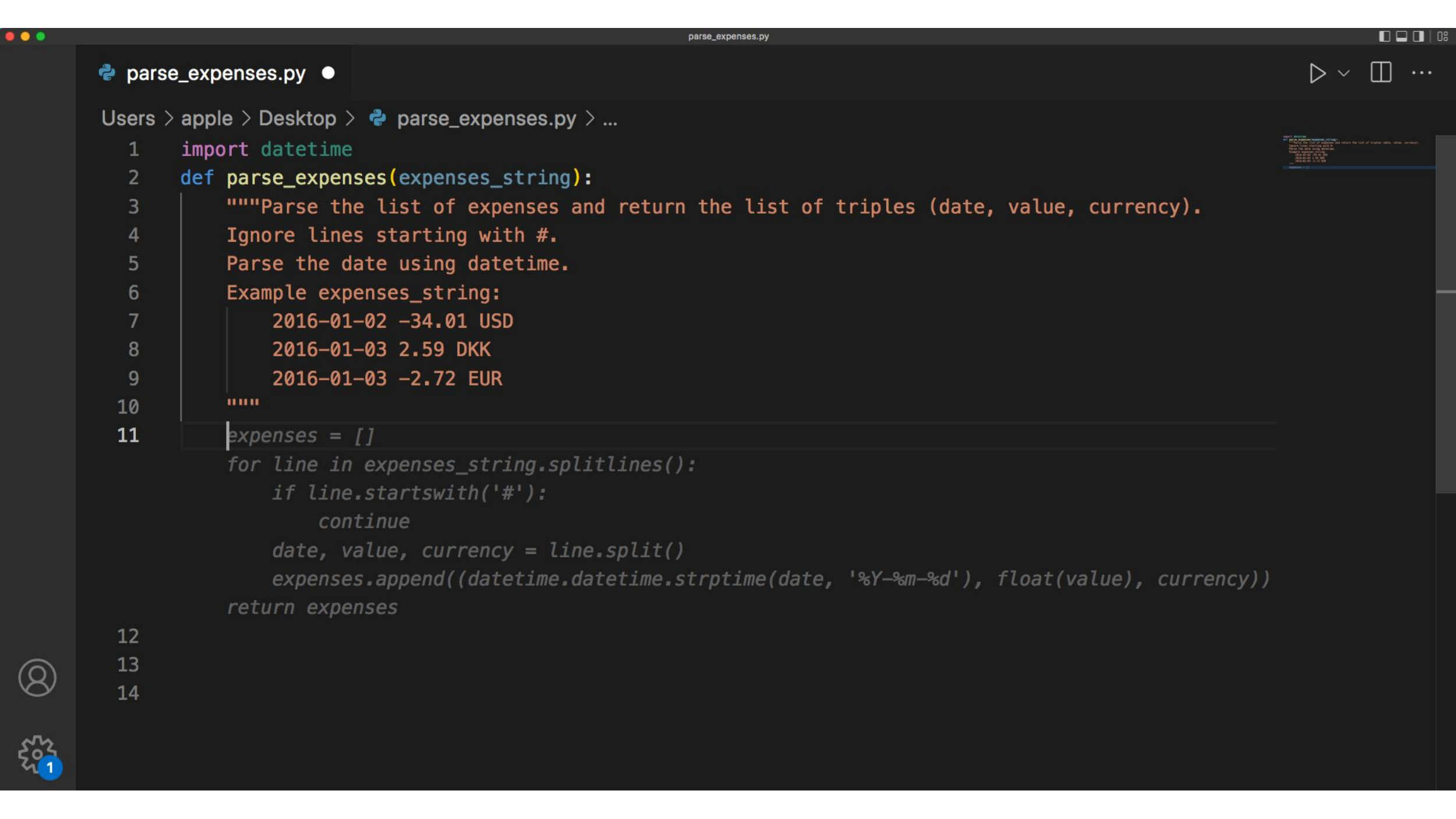}
  \caption{A screenshot of Copilot suggesting code. Copilot suggests the code snippet in the form of ghost text at the cursor location.}
  \Description{A screenshot of Visual studio code IDE with sample code prompt at the beginning of the file and code suggestion generated by copilot followed immediately after }
  \label{fig:copilot}
\end{figure*}


The recent advancement of generative AI technologies has led to revolutionary AI-powered code generation tools. Commercial AI-powered code generation tools, such as \gh Copilot (referred to as ``Copilot'' in the rest of the article), Tabnine, Kite, and Amazon Code Whisperer, and more broadly, ChatGPT, are rapidly being adopted and used by millions of developers in software development workflows \cite{Helgesson2019,Goncales2019}, influencing real-world software quality and safety. These AI tools, powered by large language models such as GPT-3 \cite{brown2020language}, can generate real-time code suggestions while the user is writing code in their programming environments. For example, Copilot, powered by the OpenAI Codex model\footnote{\url{https://openai.com/blog/openai-codex/}} trained on a large corpus of code repositories on \gh, serves as an extension in Visual Studio Code\footnote{\url{https://code.visualstudio.com/}} and can suggest code snippets in ghost text at the user's cursor location. Figure \ref{fig:copilot} presents a snapshot where Copilot is suggesting code. Specifically, Copilot can suggest code based on natural language comments (referred to as ``prompts'') and also suggest natural language comments based on user-generated code or prompts, and autocomplete single or multiple lines of code as the user types. Users can view multiple options for some of the suggestions and decide which to accept, reject, or make edits. 

These AI-powered code generation tools introduce both opportunities and risks to the programming workflow. Researchers in the HCI community only recently started to pay attention to developers' experience using AI-powered code generation tools. Both \citet{Weisz2022better} and \citet{xu2022in-ide} examined the productivity impact of software engineers working with generative code models and found opposite results: the former found that some (but not all) developers produced code of higher quality, while the latter did not find a quantitative benefit to working with the AI model. \citet{madi2023how} assessed the readability of code produced by Copilot, compared to human-written code, and found that it was comparable. A lab study investigating the usability of GitHub Copilot indicates that although developers perceive Copilot as helpful, they face difficulty assessing the correctness of the suggested code~\cite{vaithilingamExpectationVsExperience2022}. In another study, \cite{Weisz2021PerfectionNR} showed that although AI-powered code generation tools might produce problematic code, software engineers were still interested in working with it. Developers typically face little challenge in getting on board and making Copilot produce usable code; however, due to the versatility of generative AI and the resulting uncertainty in its output, developers often struggle to establish an accurate mental model of what Copilot can or cannot do and have trouble using it to its full capacity~\cite{sarkarWhatItProgram2022}. As a result of inaccurate understandings of the capacity of Copilot, developers sometimes face the challenge of making Copilot generate desired outputs~\cite{barkeGroundedCopilotHow2022}. 

These mixed signals indicate that supporting users in building appropriate trust with such tools is challenging, yet necessary. While online communities showed promises in supporting developers, it is unclear how these communities could be leveraged and enhanced to support developers' trust-building process with emerging AI-powered code generation tools. To explore the needs and opportunities in the space, we ask the following research questions:
\begin{itemize}
    \item RQ1: How do online communities help developers build appropriate trust in AI code generation tools?
    \item RQ2: How can we design to help developers build appropriate trust in AI using affordances of online communities?
\end{itemize}

\section{Study 1: How do online communities help developers build appropriate trust in AI code generation tools?}
\label{sec:study1}
To understand the role of online communities in helping developers build trust with AI code generation tools, we conducted a semi-structured interview study with 17 developers. Our findings surface two key aspects that online communities can offer to facilitate developers to build trust with AI code generation tools: \textit{community-curated experiences} and \textit{community-generated evaluation signals}. In §\ref{sec:study1_finding1} and §\ref{sec:study1_finding2}, we explain these two aspects in detail. 

\subsection{Methods}
\subsubsection{Participants and recruitment}
We recruited 17 participants for this study. Because we wanted to investigate how online communities facilitate developers' trust with AI code generation tools, we recruited participants who have experience consuming or sharing content about AI code generation tools in online communities. We posted our recruitment messages on Reddit, Twitter, and LinkedIn. All participants received a \$50 \footnote{This is an appropriate compensation for participants as it is above the minimum wage in the United States.)
} digital gift card as compensation for their participation in this study. The ethics of this study were reviewed and approved by our institute's Institutional Review Board (IRB).\footnote{anonymized for review}

We intentionally included participants with a balanced mix of demographic information, programming experience, and use of AI code generation tools and online communities. Our participants consisted of 15 men and 2 women who are between 18 and 44 years old and of a mix of multiple races. All of our participants write computer programs regularly for their jobs as software engineers, students, researchers, and IT personnel, with programming experience ranging from 3 to 16 years (median = 10 years). Our participants have used a range of AI-powered code generation tools, including Copilot, Tabnine, Kite, CodeWhisperer, and aiXcoder. They have engaged in discussions about these tools on a range of online platforms, including GitHub, Stack Overflow, Stack Exchange communities, Reddit, Hugging Face Spaces, YouTube, Twitter, and group chat channels such as Discord and Slack. Detailed profiles of our participants are presented in the Appendix (Table \ref{tab:study1-participant}).
			
\subsubsection{Procedures and analysis}
All the interviews were conducted online from July to August 2022, video recorded and later transcribed. The interviews were semi-structured. Although there is existing work that quantitatively measures user trust in technology, since our study is the first known study that strives to understand how developers \emph{form} trust in AI through participating in online communities, we chose the semi-structured interview format to keep our inquiries open-ended and to capture a breadth of themes that spoke to the experience of participants. The interview started with questions about the general experience of the participants and their thoughts about the AI code generation tools they use, how they trust the tools, and what they mean by trust. Then based on the participants' answers, the interviews dove deeply into participants' experience using online communities and how it impacts their trust, including how engagement in online communities impacts participants' expectation of AI, how and when they use the tools, and what they would do in vulnerable situations, such as when they are not sure about the AI suggestions. 

A total of 980 minutes of interview data were collected. The length of the interviews ranged from 54 minutes to 63 minutes, with an average length of 57.6 minutes.
To analyze the interview data, we performed an inductive thematic analysis~\cite{braun2006using} to identify core themes central to building trust through online communities. The first author of this paper first annotated lines of interview transcripts and took field notes on possible themes. The research team then discussed the codes, identified common themes, and collaboratively constructed a codebook. The researchers then reconducted two additional rounds of coding to iteratively merge and synthesize a final set of themes, which are presented in the following section. 

\subsection{Finding 1: Building trust with community-curated experiences}
\label{sec:study1_finding1}

For developers, an important aspect of trust in AI code generation tools is whether and how comfortable they are with integrating the tool into their programming workflow. It is common for developers to first discover and learn about specific AI tools from online platforms such as Twitter, Reddit, and \gh. Beyond initial discovery, developers also often continue to share their own tool experiences and explore others' experiences on these platforms. We refer to such collections of sharing about community members' interaction with AI as \emph{community-curated experiences}, which allow developers to form a mental model about the capacity of AI code generation tools and how they can use those tools, building trust in the process. In this section, we will discuss 1) what types of community-curated experiences that developers like to see, 2) how the community-curated experiences help developers build trust with AI tools, and 3) what challenges developers face in sharing and consuming experiences in communities. 

\subsubsection{What types of community-curated experiences are helpful.}
\label{sec:study1_finding1_needs}
Developers seek specific features in community-curated experiences that can effectively help them build and calibrate trust. We identified four important features of an idea experience sharing post: vivid description and explanation, realistic programming tasks, inclusion of diverse use cases, and details on setup and dependency.

\paragraph{Vivid description and explanation.}
One of the most important features that developers would like to see in community is others' specific experience interacting with an AI---in P13's words, ``concrete examples of how someone's using a tool and how it helps them.'' These posts typically contain details including the programming tasks, the prompt that elicits AI suggestions, and the outcomes of the AI suggestion, often supported by screenshots or video recordings. For example, P2 shared that they were convinced to try out Copilot by a YouTube video because they were able to see ``somebody actively working with it and I can be like, that's how it would work for me too.'' (P2) Similarly, P13 stated that a video demo could vividly display how the AI assisted users and help them recognize the benefit it could bring: ``when you watch it, parts of it can translate on whatever you do and you can make your workflow better.'' In addition to a vivid, detailed description of the interaction, developers also would like to see the users' explanation, opinions, and reflection on the interaction and AI suggestion. For instance, P3 explained that an effective demonstration of a vivid experience with AI should explain that ``I wrote this suggestion---look what it was able to come up with, and why you should use it as well, because it could help you with something like this.'' With the explanation, developers can achieve a more accurate understanding of how AI behaved in the user example.

\paragraph{Realistic programming tasks.}
It is also important that the shared experience is situated in a programming task that represents the typical and realistic workflow of a developer. Instead of using AI to solve predefined programming challenges, developers want to see the user showcase an ``actual project'' where they are ``deeply nested into'' and working with ``lots of files, lots of context all around.'' (P3) This type of demonstration can help developer anticipate the performance of AI in complex situations that they face in their day-to-day work. 
Specifically, developers would like to see both cases where AI makes helpful suggestions and fails to help. For example, P16 preferred to see posts without ``a pre-biased judgment'' and stated that ``it's better to see how people who are actually using it in real world face issues.'' Many agreed that it was important to see incidents where the AI gave wrongful suggestions so that they could understand its limitation. Working with AI on a realistic project gives opportunities for such incidents to occur:

\begin{quote}
``I think the intention shouldn't be to use Copilot or not. They can just keep it running in the back and look at its suggestions and just work out they would normally work... Then you can see if it helps, it helps, but otherwise you're just dismissing the suggestions. Then we would know whether it is helping.'' (P10)
\end{quote}

\paragraph{Inclusion of diverse use cases.}
\label{sec:diverse_use_case}
Furthermore, developers would search for diverse use cases of an AI tool---different programming languages, tasks, scenarios, and so on---to learn what is possible with its capacity. For example, P13 shared their experience going down in a long ``YouTube hole'' when first learned about Copilot and got exposed to the breadth of its use cases: ``there's so much capability it has! I had a lot of fun just watching the way people use Copilot, front-end development, back-end development, machine learning development...'' When introducing Copilot to their friend, P13 always demonstrated it in a variety of programming languages and tasks, including how it successfully suggested complex function in OCaml but stuck with the ``<div>''s in HTML: ``Because that's what I experienced, I'm going to show them the full picture of what I've done.'' (P13)  Diverse use cases also help developers be aware of the boundary of capacity of the AI. Similarly, P2 shared that they once posted about how they used Copilot in a multi-component, full-stack web development project Copilot to demonstrate the ability of Copilot. P2 was proud of this post, as it showed how Copilot behaved in a variety of programming languages and tasks, providing a comprehensive picture of its capacity: 

\begin{quote}
``I think just like one specific code example isn't really enough\dots{} If one person says, here's a screenshot of a cool thing co-pilot recommended, I don't know if that be enough to convince me\dots{} That doesn't prove that like co-pilot is good overall.'' (P2)
\end{quote}

\paragraph{Details on project context and dependency.}
\label{sec:dependency}
In order to make use of the experience of others with AI code generation tools, developers strongly favor posts that include details on setup and dependencies of the project, for example, ``the tool, what [the user] was planning to program, the language, the purpose, everything.'' (P1)
Developers value information on setup and dependency because they often want to reproduce the interaction with AI in their use cases. For example, P1 preferred reviews that are ``Something someone can follow and you'll get the expected results.'' In another example, P6 once saw a video about a similar problem that he had with Copilot and would like ``see someone doing it, just follow it step by step and to see.'' Replicability is crucial to trust building, as our participants explained that they needed to get hands-on experience with an AI tool before deciding how to trust it. Developers would like to replicate others' successful interaction with AI to help with their particular use cases, while sometimes also examine wrongful suggestions that they saw online before making their own judgement: ``If I see something negative I don't like to say that, okay, this is all useless let's stop using Copilot\dots{} I try it with caution, I try to reproduce it.'' (P4)

\subsubsection{How community-curated experiences help developers build trust with AI tools.}
Our finding suggests that developers can build appropriate trust with AI code generation tools as a result of engaging with community-curated experience. Specifically, developers can build a mental model of the capacity of the AI from others' experience, which includes setting reasonable expectations, learning strategies on when to trust the AI, forming empirical understanding of how AI generates suggestions, and developing awareness of broader implications of AI-generated code. \\

\paragraph{Setting reasonable expectations on the capability of the AI}
Engaging with community-generated content on experience with AI code generation tools helps developers form an appropriate level of expectations with those tools. Before trying out the tool, it is common for developers to have little knowledge about its ability and how it will perform in their particular use cases, as P7 put, ``I've not tried anything like that before so I didn't really know what I was expecting.'' Consuming other users' ``anecdotes'' with the tools allows developers to form an initial mental model about the AI and decide if it is trustworthy enough to give it a try: ``I read a bunch of what people think of the eventual outcome... [It] helps me make my own perception of whether it is something that is useful for me or not.'' (p16)

Forming reasonable expectations also means learning about important constraints of AI. An unrealistically high expectation could undermine developers' trust, as they may not be prepared for its constraints. For instance, P14 shared that they initially thought Tabnine could largely automate their programming tasks due to an online post claiming that Tabnine wrote 20-30 percentage code in a file, but then found that it was not the case in their actual usage: ``I thought it was going to give me more detailed prediction on my code\dots{} But once I tried it out, I just saw that it`s mostly just complete some lines.'' (P14) Out of disappointment, P14 stopped using Tabnine. In contrast, a reasonable expectation allows developers to anticipate certain failures, so that their trust would not be broken by surprising negative experiences. For instance, P13 shared that a video demonstrating a variety of successful and failed use cases of Kite helped them realistically estimate how it can improve productivity: ``I set my expectations lower. I wasn't like, `wow, I thought Kite would write my entire code file for me.' '' Such realistic expectations allow developers to understand the boundary of the AI's capacity and plan out how they will use the AI accordingly. For example, P4 anticipated that Copilot would ``fail in some cases'' and did not expect Copilot ``to work all the time'', but still decided to try it out with caution and validation. 

\paragraph{Learning strategies of when to trust the AI}
Community-generated content can also allow developers to develop strategies on how to trust and effectively use AI in their specific use cases. While many are unsure whether they could trust AI for particular tasks in their workflow, developers go to online communities to seek relevant examples. For instance, P16 shared that they looked into community discussions to learn the tasks that they could trust Copilot to do:
\begin{quote}
``The public discussion has definitely helped with the trust: this is a language translation that I can trust. This is a module transmission I can trust\dots{} It's like all I need to do is find already if someone has tried to use the auto-complete.'' (P16)
\end{quote}

Developers can also learn from others' experience about areas where AI should not be trusted, and consequently form strategies to minimize any harm. For example, P1 shared an experience in which they tried to get Kite to make high-quality suggestions in a project that involved multiple programming languages. After browsing in online communities, they learned from other users that Kite generally did not perform well in some languages and decided that they should instead ``stick to a particular programming language.'' In particular, developers would like to see examples relevant to their specific use cases, as they offer previews of how the AI can function in their use cases, allowing developers to decide how much to trust it and whether to adopt it at all: 
\begin{quote}
``When I found out the reviews does not affect the areas of function that I want to use the app for, I will not really be bothered about checking out the app. But when I found out the negative reviews is [in] the area where I want to use the application for, I tends to not have enough trust for the application and mostly I end up not checking out at all.'' (P5) 
\end{quote}

\paragraph{Forming empirical understandings of how the AI generates particular suggestions.}
Online communities can also help developers to understand how AI code generation tools work. Developers often contemplate about why the AI gave certain suggestions in particular scenarios, for example, ``why does it make a certain choice over another? Why is the naming convention of this variable something else?'' (P8) Knowing the inner decision processes and rationales of AI is crucial to trust, as users can make decisions based on the reliability of the decision process. In online communities, members share their prompts and AI suggestions, and collectively build an empirical understanding for the otherwise black box:``it's a super black box, so we're collectively trying to figure out how to best use them.'' (P10) In particular, participants shared appreciation for discussions around wrongful AI suggestions. In those discussions, users post bizarre code written by the AI and the surrounding context, speculate potential causes of those suggestions, and sometimes exchange ideas for experiment to test out their theories. For example, P4 shared an experience where they engaged deeply in a discussion thread and tried to reverse engineer the prompt that resulted in a wrongful suggestion by Copilot: ``I'd like to see more of this type of discussions. [We] see examples where it is going wrong and trying to figure out what went wrong.'' (P4) Similarly, P16 enjoyed reading posts in which users shared experiments with GPT-3 prompts, as those discussions helped them empirically summarize the factors that affect AI suggestions: 
 
\begin{quote}
``If you see the problem that they wrote and you see the paragraph that they get then it's very easy to figure out the breaking points that this word is contributing for this sentence to be generated, and basically bring a bit of interpretability to the machine learning model.'' (P16)
\end{quote}

\paragraph{Developing awareness of broader implications of AI generated code.}
Developers also learn about the implications of using AI models beyond their immediate use cases by participating in online communities. These implications can include the legal and ethical impact of the code generated by AI, as well as potential concerns about security. For example, P3, a young programmer, only learned about the controversial impact of Copilot on FLOSS\footnote{\url{https://www.gnu.org/philosophy/floss-and-foss.en.html}} after following a debate on GitHub repositories:

\begin{quote}
``If I had not seen the discussion, I don't think I ever would have thought about it\dots{} But once I did see a lot of people talking about the licensing issues\dots{} I feel like I should be a little more careful with what I do.'' (P3) 
\end{quote}

Although those legal and ethical implications would not directly affect P3's current use cases of Copilot, P3 was glad to be aware of those issues to prepare for wider use cases in the future. In terms of security implications, while experienced developers may have knowledge of the vulnerabilities that AI can introduce, beginner programmers rely on community-generated resources to learn about these concerns. For example, participants reported that they had seen posts about Copilot suggesting others' private keys. These implications are crucial for trust, as they can help developers stay alert and stay away from serious consequences: 
\begin{quote}
``People learned that they should be maybe careful against this in critical software. That's a good thing that people were discussing it because there's probably less probability of the worst-case happening.'' (P12)
\end{quote} 

Seeing security issues introduced by AI reminds developers of the importance of manual review on AI suggested code and their role as the supervisor of the AI. For example, P8 recalled that a discussion about a security breach caused by code written by Copilot made them invest more time in verification: 

\begin{quote} 
``I would be more careful while using the application, just so I would not just write all my code and just publish it that way without taking time to actually verify.'' (P8)
\end{quote}

\subsubsection{What challenges developers face in sharing and consuming experience in communities.}
\label{sec:study1_finding1_challenge}
Despite its benefits, developers face a variety of challenges in engaging with community-curated experience. Some participants found it is hard to reproduce others' experience since vivid description and dependency details were often missing, and others complained about the lack of diverse use cases or realistic programming tasks. 
We summarized three significant challenges that developers face in effectively sharing or consuming experience with AI tools: lack of channels dedicated to experience sharing, high cost of describing and sharing experience, and missing details for reproducibility.

\paragraph{Lack of channels dedicated to experience sharing.}
A common challenge developers face is the lack of platforms dedicated to sharing experience with AI code generation tools. On general platforms like Twitter and Reddit, sharing and discussions about specific experience with AI code generation tools can be buried in more general discussions about AI. For example, 
P16 shared that they had been apprehensive about engaging in high-level discussions about AI, since those discussions could often become controversial and distant from specific user experience:

\begin{quote}
``Most of the discussion start with some random Atlanta or New York post headline saying something like, AI is going to take your job or anything like that\dots{} the headline is so sensational that people have really strong opinions over it and I do not really like engaging with.'' (P16)
\end{quote}

The participants expressed a desire for a channel dedicated to end-user experience sharing and discussions about AI code generation tools. As P10 imagined, an example of the channel could be ``a common forum just specific for Copilot or something, like crowd sourcing of ideas.'' Similarly, P13 suggested a ``centralized place'' for Copilot related experience, where people share ``code snippets and they'll be, `here once you have access to Copilot, try this code snippet in your repository, see what happens.' '' In addition, P7 proposed that a search function for ''the feature you want and bring out posts about related AI co-generation tools'' so that users can easily locate examples with particular use cases.

\paragraph{High cost of describing and sharing experience.}
For many developers, writing a post about their experience with AI code generation tools and sharing it online is considered a time-consuming and complex task. Given the interactive nature of AI tools, it is difficult to effectively describe the interaction process. For example, P2 shared that it is hard to describe AI suggestions when making a text-based social media post: ``You can't really have a code block that differentiates between I wrote this code and then this is the code that was recommended.'' (P2)
For developers who choose to share a video about their interaction with the AI, make the video consumable by other developers requires a lot of planning on what to record and how to explain. For example, P2, who once prepared an elaborated video post about how Copilot helped them in a complicated project (see §\ref{sec:diverse_use_case}), complained about the tedious process of preparing the video: 

\begin{quote}
``I spent a lot of time learning the intricacies of, like, for that specific project, what would Copilot recommend and all that stuff. I was up until like 7 a.m.'' (P2)
\end{quote}

In addition, posting on a public online platform can be intimidating. Developers may feel self-conscious when sharing their experience and opinions to a large audience, with little knowledge about who they are and how they would react. As a result, developers may want to spend a lot of time polishing the post that they are trying to share, adding more investment to this task. P11 compared sharing their experience with Copilot to a group of friends with making a public post:

\begin{quote}
``(With friends,) I don't really care about going deep into discussion in order to appear perfect. But when it come to posting online, I'm most mindful of how my writing will look.'' (P11)
\end{quote}

\paragraph{Missing details for reproducibility.} Another common challenge is the lack of reproducibility in the experience shared by developers online. As we show in §\ref{sec:dependency}, developers build trust with AI by trying out prompts from online sources in their workflow. However, when sharing their experience, developers often overlook the setup of their project and the dependency of their environment: ``people don't exactly share their VSCode settings, [which] could be very much catered to the way they code or their programming preferences.'' (P9) Developers are disappointed when they are unable to replicate AI interactions that they saw online in their own environment. 
For example, P13 complained of the lack of functionality to ``copy and paste texts from the video'' (where the other user interacts with Copilot) and wished a``testing environment'' where they could directly replicate the interaction.
The lack of reproducibility can make developers not aware of diverse use cases and possible limitation of AI, resulting in biased views. Some of our participants even deemed particular use cases that they saw online but had not experienced themselves as ``misinformation'', because they were not able to confirm themselves that such interaction is possible: ``if I see that people are contrary on something that can't be verified, I would be like, this is a misinformation.'' (P7)

\subsection{Finding 2: building trust with community-generated evaluation signals}
\label{sec:study1_finding2}
Another important aspect of trust in AI code generation tools for developers is whether and how they can leverage specific AI suggestions. Developers can find signals to evaluate the code suggested by AI from code solutions posted by community members and make decisions. We refer to these contributions of community members as \emph{community-generated evaluation signals}. In the following sections, we explain how and why communities can offer effective support for the evaluation of AI output. Specifically, we unpack 1) what evaluation signals communities can offer, 2) how evaluation signals help developers build trust with AI tools, and 3) the trade-off between evaluation signals and productivity.

\subsubsection{What evaluation signals communities can offer.}
\label{sec:study1_finding2_need}
We identified three major evaluation signals that developers' online communities can offer: direct indicators of code quality, context and generation process of a code solution, and identity signals.

\paragraph{Direct indicators of code quality.}
Unlike AI code generation tools, online communities offer direct metrics that can help developers evaluate the quality of code solutions. Many online communities, such as Stack Overflow, allow users to vote or provide ratings for solutions. Participants trusted a solution selected by such voting mechanism as it had been reviewed by many others: ``if other programmers have used that solution and it worked in their code, they'll upvote the solution and so that does give you a little bit more faith that this is a good solution.'' (P3) 
The voting mechanism also implies that multiple real developers have used the code, indicating the safety and trustworthiness of the code. For instance, for P10, the solution selected by a question owner on Stack Overflow indicated that ``the person who originally asked the question has tried it out and it worked for them,'' providing ``some guarantee that, that code compiles or it's very close to what I want, which does not exist with Copilot.'' 

Another explicit indicator of code quality is how much engagement it receives. In online communities, developers can see how many other members viewed, rated, shared, commented on, or otherwise engaged with a post. High and positive engagement with a solution usually means that it ``has been cross-verified by a lot of people,'' (P4) indicating high quality. As P6 summarized, ``when it's coming from plenty of different people and all bring positive reply, I'll be like, `yeah, it works for everyone and not just for a single person.' '' With high engagement, developers can see verification triangulated from multiple sources, which informs their evaluation: ``I trust that because it is not dealing with just from one person's perspective---different developers with different ideas, coming together to make their review.'' (P1)  When a solution does not get high engagement, developers tend to be more skeptical about it since it is not verified by other users. For example, P14 explained the reason that they did not have a high degree of trust with Tabnine after seeing a single post about it: ``because of the fact that just a random post from a random user with no really much engagements.'' (P14)

\paragraph{Context and generation process of a code solution.}
Developers also appreciate that online communities offer a broader context beyond the code solution itself that helps with evaluation. Platforms such as Stack Overflow support discussions around solutions contributed by users. Participants expressed appreciation to the discussions, as compared to solutions suggested by AI, they are more ``detailed'', ``interactive'' (P12), and provide ``a lot more transparency'' to how it is generated (P3). Following a discussion allows developers to understand why a solution is chosen as the optimal: 

\begin{quote}
``They will go through the process. What happened? How did it happen? \dots{} then people will go through different solutions. There is a lot more storytelling involved.'' (P9)
\end{quote}

\begin{quote}
``You can see this back-and-forth between people until one answer becomes the accepted answer for that question. Somebody is suggesting some code and then in the comments people point out, oh, you have a point here or like there's an issue with this thing.'' (P12)
\end{quote}

Understanding how a code solution is generated helps developers better judge its quality. Because users cannot see how an AI suggestion is decided, they have trouble deciding its validity, as P3 described: ``in some cases you have no idea why something is working, although it works.'' Sometimes, the solution generated by AI may not be the most optimal solution. For example, P3 shared an experience where Copilot suggested a bug-free, but inefficient implementation of a function. While individual inexperienced programmers often have trouble identifying issues in examples like this, they can leverage online discussions where multiple, more experienced members can scrutinize the solution, spot any issues, and even propose more effective alternatives, as P3 reflected: ``if [I] had gone with Stack Overflow for something like that, they would have said, `here's the inefficient solution, this is why it's inefficient, and here's how you can make it more efficient.' ''

\paragraph{Identity signal.}
Participants agreed that knowing the identity of the author of the code allows them to evaluate the code based on the authors' background and experience level.
In communities like Stack Overflow and \gh, users can view members' contribution history indicating what and how much code they have contributed and how well the code has been received in the community. Common gamification mechanisms such as levels and badges also signal the expertise of a user. To developers, all these factors can help them decide to what extent they can trust the code written by a particular user: 

\begin{quote}
``I'm copying this code of this guy which has all the badges and everything---he probably knows what he's doing.'' (P12)

``On Stack Overflow they have a reputation. You see they answered 3000 questions and they're always posting high quality responses, and then that's a pretty big proponent that you can trust that person\dots{} then you can most likely trust their code.'' (P3)
\end{quote}

Especially when the developer is unfamiliar about an area or indecisive about multiple options, an expert's input can facilitate their decision process. For instance, P15 described a scenario where signals of expertise could help them recognize important considerations: 

\begin{quote}
``It may be the case that thousand people voted for it, but two or three people may have commented saying, `Hey, it has a threat.' And those two people who commented are actually threat analysis specialists. It's not always 1,000 people's opinion are correct.'' (P15)
\end{quote}

\subsubsection{How evaluation signals help developers build trust with AI tools.}
Currently AI code generation tools lack a evaluation mechanism for the quality and correctness of the AI-suggested code, as P3 observed with the example of Copilot, ``GitHub Copilot just hands you the code, and it's up to you to know whether or not that's good code.'' While developers can evaluate the suggestions themselves by reading or executing the code, external support for evaluation is often needed when cost of underlying error is too high in terms of computing resources and time. Especially for developers with less experience who may be unfamiliar with the syntax or logistics of the suggested code, effective support for evaluation is essential to ensure quality and safety in the code, as P7 summarized: ``there are times that it's giving me suggestion I'm not very familiar with and I would have to look it up and see.'' Online communities can also provide multiple perspectives to triangulate evaluation: ``I trust that more because it is not dealing with just from one person's perspective. It deals with different programmers coming together with different ideas.'' (P1) In addition, knowing that online communities attract users of a variety of expertise and experience levels can boost developers' confidence for evaluation. As P7 pointed out, ``online community has lots of people with more knowledge and who has worked on more projects probably than I have, so they're familiar with many codes that I'm not,'' developers can rely on online communities to learn more about the code and make appropriate judgements. 

\subsubsection{Trade-off between evaluation signals and productivity.}
\label{sec:study1_finding2_challenge}
Participants pointed out an important trade-off between the effective evaluation that a community can provide and the amount of effort that a user needs to invest in a community. A great advantage of AI code generation tools is that they increase productivity---developers get AI suggestions seamlessly in their programming workflow, reducing the time spent typing and looking up online for syntax and documentation. As P16 framed, Copilot is ``doing the filtering out process that I need in Stack Overflow to do myself for me\dots{} It helps me not open 50 Google Chrome tabs and still have the answer in an efficient amount of time possible.'' Whereas in online communities like Stack Overflow, they need to ``look through multiple questions and look for multiple answers and figure out whether this person wants to do the exact thing that I want to do.'' (P16) In other words, it can take an enormous amount of time and energy to search, filter, and collect community generated resources and use credibility signals to validate those user sharing. This can break the workflow of a developer since their tasks are often very targeted and time very constrained. As P15 stated, ``software engineers, they're not go and search for online contents, they work on a need to do basis.'' There is the need to effectively incorporate community evaluation in developer's programming workflow.

\section{Study 2: How can we design to help developers build appropriate trust in AI using affordances of online communities?}
\label{sec:study2}
While Study 1 showcases how community-curated experiences and community-generated evaluation signals help developers build appropriate trust with AI code generation tools and what developers seek in these types of information, we also identified many challenges and needs that developers had when forming an appropriate level of trust. To further explore the design space of incorporating a \textit{user community} into the experience of AI code generation tools and helping developers form appropriate trust, we conducted Study 2, a design probe study with 11 developers. 

Specifically, we first developed two sets of design concepts---\textit{community evaluation signals} and \text{experience sharing spaces}---in the form of visual stimuli. We presented these visual stimuli as \emph{design probes} to a new set of participants in imaginary use cases to probe for their reaction and any further ideas to improve the designs. Not measured by quantitative metrics, this type of qualitative design probe study is common in formative research on human-AI interaction (e.g., \cite{sunInvestigatingExplainabilityGenerative2022a, cheng-etal-2022-mapping}), as our goal is not to settle on specific design elements, but rather to explore what is possible in the design space. Probed by our visual stimuli, participants were also able to generate new ideas and suggestions. The output of this section is a series of elaborated design directions that we created based on feedback from our participants.

In the following sections, we list out the user needs identified from Study 1 interviews, explain how we develop the design probe stimuli and the rationales for specific features, and dive into the findings on how the two sets of stimuli can help developers build appropriate trust respectively.

\subsection{Identifying User Needs}
\label{sec:study2_user_needs}

\paragraph{Community-generated evaluation signals.} 
Based on our findings presented in §\ref{sec:study1_finding2}, developers tend to seek community-generated evaluation signals to judge how much they can trust AI code suggestions. We summarize the following user needs and challenges in leveraging evaluation signals to build trust: 

\begin{enumerate}
    \item Users seek direct indicators of code quality in communities.
    \item Users appreciate community discussions that allow them to learn more about the context of a code solution.
    \item Users would like to see identity signals of community members to decide the credibility of their contributions.
    \item Users feel distracted when having to leave the development environment in the midst of a coding session to search for code suggestions on external platforms.
\end{enumerate}

\paragraph{Community-curated experiences.} 
Our findings in §\ref{sec:study1_finding1} indicate the importance of community-curated experience to help users build trust with AI code generation tools. We summarize the following user needs and challenges in consuming community-curated experiences and sharing experience with the community: 
\begin{enumerate}
    \item Users would like to see others' specific experience interacting with AI tools, shared with vivid description of interaction and situated in realistic programming tasks 
    \item Users want to explore diverse use cases of the AI tool from community-curated experiences
    \item Users would like to replicate certain interaction with AI that they learned online and need information on dependency and project context
    \item Users cannot find a dedicated channel for developers to share specific experience with AI code generation tools.
    \item It takes a high cost in time and effort to document and share experience.
\end{enumerate}

\subsection{Building Design Probe Stimuli}
\label{sec:study2_design}
\begin{figure*}[t]
     \centering
     \begin{subfigure}[b]{0.48\textwidth}
         \centering
         \includegraphics[width=\textwidth]{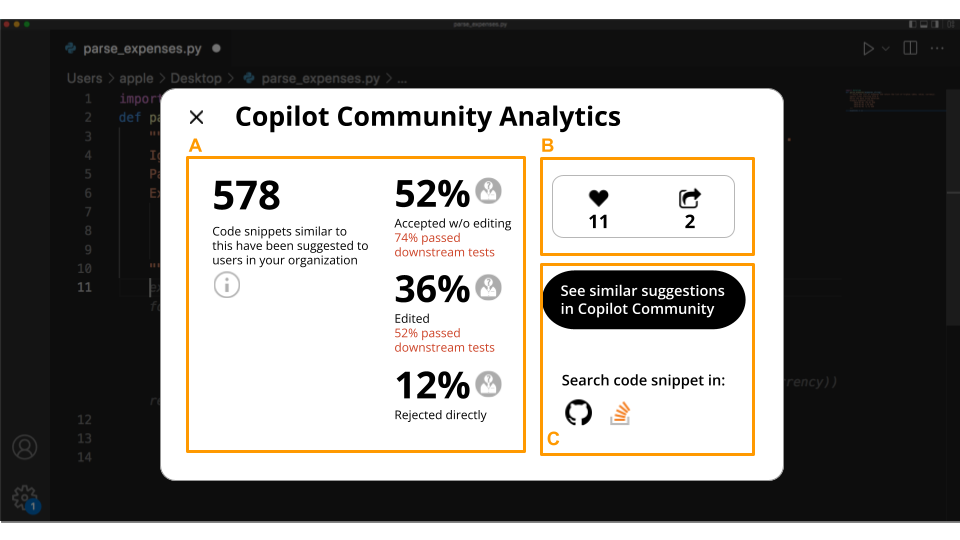}
     \end{subfigure}
     \hfill
     \begin{subfigure}[b]{0.48\textwidth}
         \centering
         \includegraphics[width=\textwidth]{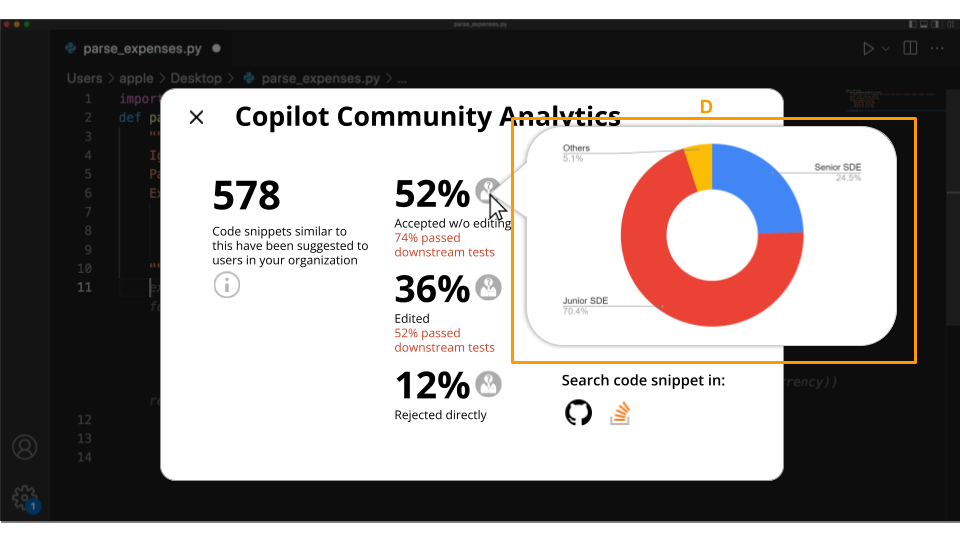}
     \end{subfigure}
     \caption{Community Evaluation Signal Stimuli used in Study 2. (A) Community usage statistics; (B) Community voting; (C) Link to external communities such as Stack Overflow and \gh; (D) Percentage breakdown of community usage statistics by job titles. }
     \Description{Left image showing a popup dashboard titled Copilot Community Analytics. The dashboard shows that 578 Code snippets similar to this have been suggested to users in your organization; 52\% Accepted w/o editing
74\% passed downstream tests; 36\% Edited
52\% passed downstream tests; 12\% rejected directly. Users can click on a button says See similiar suggestions in Copilot community. The dashboard also have icons showing number of likes received from community member, and \gh and Stack Overflow icons where users can search code snippets in these communities. }
     \label{fig:community_signal}
\end{figure*}

\begin{figure*}[t]
     \centering
     \begin{subfigure}[b]{0.48\textwidth}
         \centering
         \includegraphics[width=\textwidth]{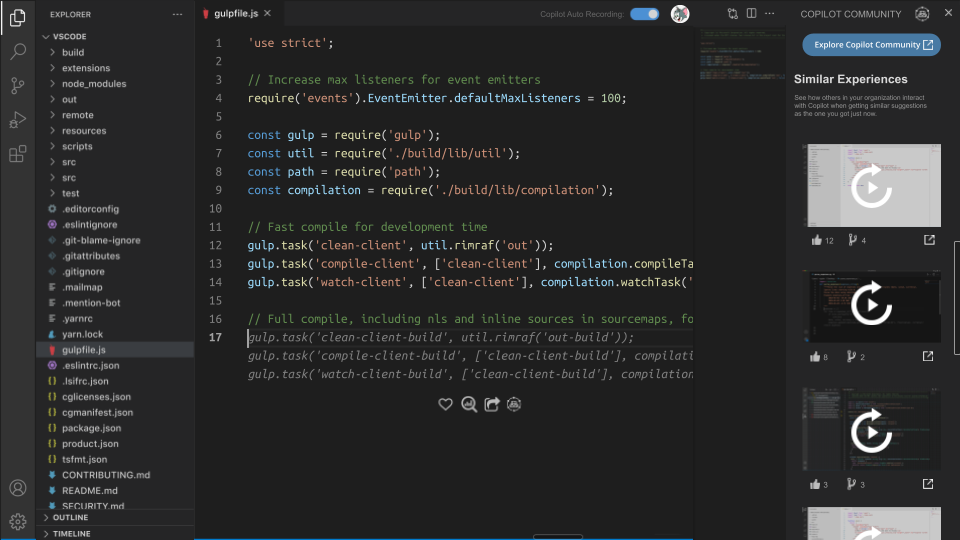}
         \caption{Experience sharing space within IDE }
         \Description{A coding interface in IDE. The sidebar shows a list of video previews with their corresponding number of likes and forks. }
         \label{fig:design_repo_sidebar}
     \end{subfigure}
     \hfill
     \begin{subfigure}[b]{0.48\textwidth}
         \centering
         \includegraphics[width=\textwidth]{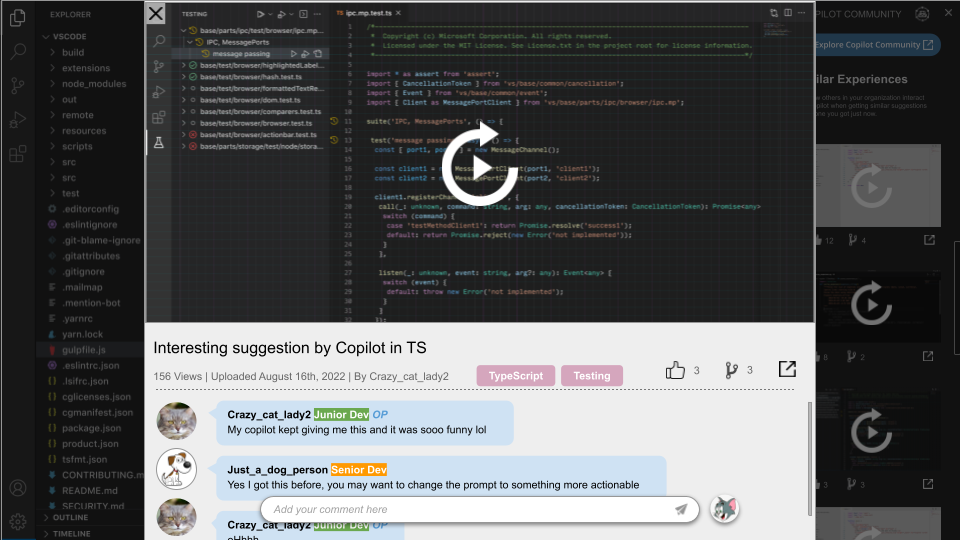}
         \caption{Example
of a experience sharing post}
         \Description{A post containing a video snippet, user generated title and labels, as well as social features such as likes and comments.}
         \label{fig:design_repo_video}
     \end{subfigure}
     \newline
      \centering
     \begin{subfigure}[b]{0.48\textwidth}
         \centering
         \includegraphics[width=\textwidth]{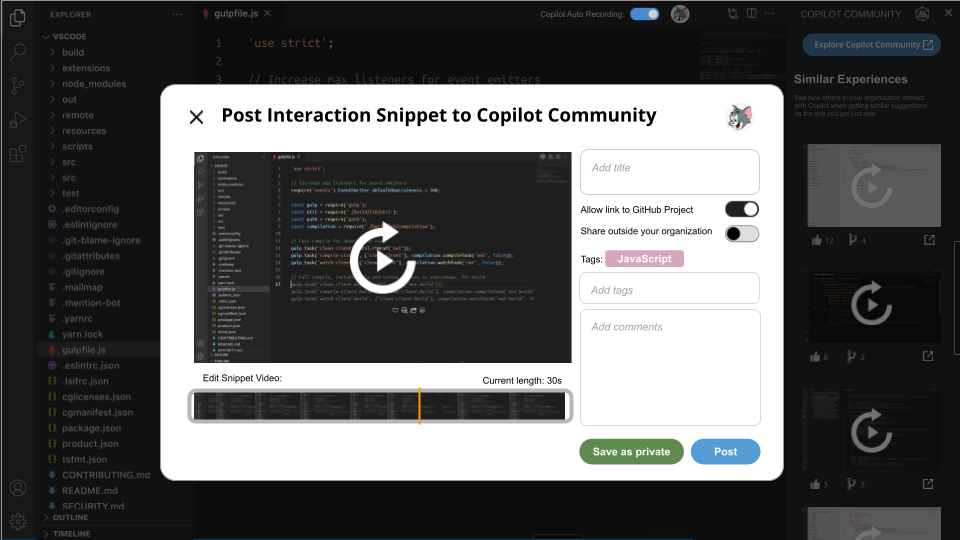}
         \caption{Post sharing and editing interface}
         \Description{A post editing interface including preview of a coding video, a selection bar for editing the video, and text fields to add title and description of the post.}
         \label{fig:design_repo_sharing}
     \end{subfigure}
     \hfill
     \begin{subfigure}[b]{0.48\textwidth}
         \centering
         \includegraphics[width=\textwidth]{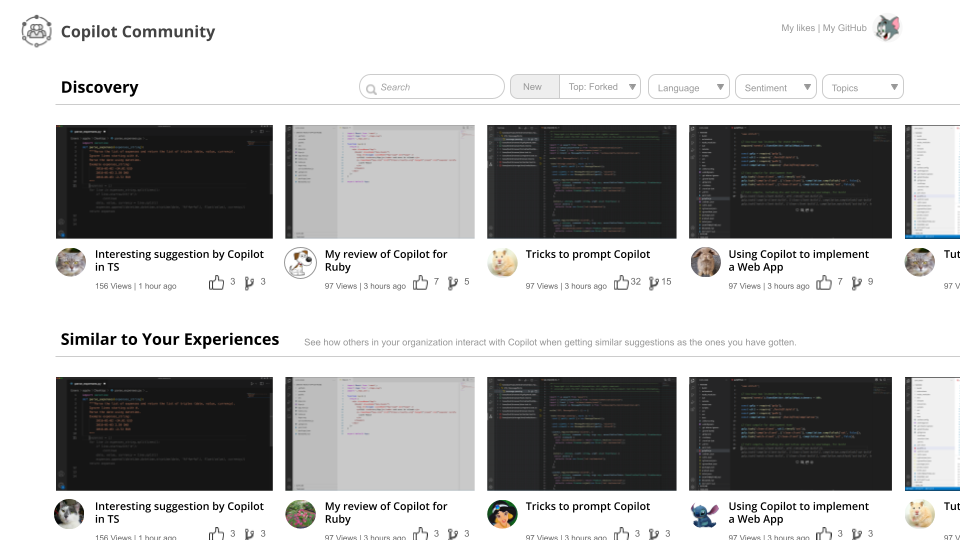}
         \caption{External experience sharing space}
         \Description{Copilot community page with two sections, Discovery and Similiar to Your Experiences. Each post shows the title, upload time, number of likes and forks.}
         \label{fig:design_repo_external}
     \end{subfigure}
     \caption{Experience Sharing Spaces Stimuli used in Study 2}
     \label{fig:design_repo}
\end{figure*}

Given the user needs presented in §\ref{sec:study2_user_needs}, we designed two sets of stimuli---the \textit{community evaluation signals} and the \textit{experience sharing spaces}---as examples to address user needs regarding community-generated evaluation signals and community-curated experiences respectively.

\paragraph{Community evaluation signals.}
The stimuli for the community evaluation signals are shown in Figure \ref{fig:community_signal} and the features are labeled as A - D. We designed an experience where users can view how other users interact with Copilot suggestions without the need to switch to an external platform. When the user gets a code suggestion from Copilot, they have the option to click on a button to view community evaluation signals in a pop-up window. We introduced feature A \textit{Community usage statistics} and feature B \textit{community voting} as indicators of code quality that allow users to see the prevalence of the suggestion used by other users, the percentage of acceptance, edits, rejection, and passing of downstream tests, as well as votings in terms of likes and shares. Feature C enables users to search code in external communities such as Stack Overflow and \gh, as well as experience sharing spaces discussed in the next section. With feature D \textit{User identity}, users can hover on a community usage statistic to see the percentage breakdown of users who took this action with respect to their job titles. \\

\paragraph{Experience sharing spaces.}
The stimuli experience sharing spaces are shown in Figure \ref{fig:design_repo}.
Since we were curious about developers' needs on community-curated experience within and outside their programming workflow, we designed two versions of dedicated space---an \textit{experience sharing space within IDE} (Figure \ref{fig:design_repo_sidebar}) and an \textit{external experience sharing space} (Figure \ref{fig:design_repo_external}). 
In both spaces, the user can view posts of short video snippets that record community members' interaction with particular Copilot suggestions. 
The space within IDE takes the format of a sidebar in the IDE window, that users can toggle once they got a code suggestion and contains community-curated video snippets relevant to the suggestion. 
The external space is a standalone platform, where users have the opportunity to explore diverse use cases using the search bar and a variety of filters. The video recording of a post speaks to the user needs of vivid description of interaction. Figure \ref{fig:design_repo_video} shows the details of a post, including a video snippet, user-generated title and labels, as well as social features such as likes and comments. Figure \ref{fig:design_repo_sharing} illustrates how a post is generated. Considering the challenge that it takes a high cost to make a post, we designed an automatic video recording feature that captures a certain amount of time before and after a Copilot suggestion to minimize user effort. In situ recording of a coding session also speaks to the user need to see realistic programming tasks. 
Besides the video snippet, users have the option to add text-based title, description, and labels to their post, as well as link the video to the code repository of the project on GitHub, so that others could see their project context, dependency, setup, and fork the project if they would like to fully replicate the interaction.

For all stimuli, we used Copilot as an example of AI code generation tools and Visual Studio Code\footnote{\url{https://code.visualstudio.com/}}---the integrated development environment (IDE) that hosts Copilot---as an example of a development environment. 
All stimuli were developed using Microsoft PowerPoint and served as elicitation to help users imagine the interaction in the design probe sessions. 
Importantly, we do not intend to propose comprehensive designs that address all the user needs that we discovered in Study 1. We also do not claim that the designs presented in the stimuli are the best solution to the problems. The stimuli are intended to represent the set of design concepts that tackle the user needs derived from Study 1 that we are interested in exploring. We used the stimuli to elicit user experience, needs, and feedback.

\subsection{Participants and Procedures}
We recruited 11 developers, consisting of 10 men and 1 woman of a mix of demographic profiles in race, age, and level of education, as well as programming experience. Because our designs were based on \gh Copilot, we intentionally recruited users who have used \gh Copilot. Participants were recruited on Twitter and LinkedIn, as well as in a large tech company. We intentionally recruited a balance of participants who work in big tech companies and small organizations and who write production code and end-user code to capture different use cases of Copilot and user needs. The detailed profile of the participants can be found in the Appendix Table \ref{tab:study2-participant}.  

All design probe sessions were conducted over recorded video calls in August 2022. In each session, we present the visual stimuli, explain each feature, and investigate the experience and needs of the participants. We encouraged participants to imagine using the features in their actual programming workflow with Copilot and speculate potential advantages and pitfalls. We also ask participants to brainstorm any changes or new features that they want to add. The ethics of this study were reviewed and approved by our institute's Institutional Review Board (IRB).

A total of 646 minutes of video recorded data were collected. The length of the sessions ranged from 54 minutes to 65 minutes, with an average length of 58.7 minutes. The analysis of the data from the design probe sessions followed the procedure of thematic analysis~\cite{braun2006using}. The first author of the paper coded the data based on the features in the stimuli and the participants' reactions and suggestions. The first author then discussed the codes with the research team and iterated on the codes. 

\subsection{Finding 1: User experience and needs with community evaluation signals} 
Participants agreed that the signals of how other users interact with the suggestion are generally useful to help them evaluate Copilot's suggestions. In the words of DP-P1, the signals can help ``accelerate the checking process'' (DP-P1). In particular, participants saw themselves referring to these signals when they were not familiar with the language or programming tasks: ``[If] it's the first time [writing the code] and I'm just not too sure about myself... I just want to see how people are using this.'' (DP-P9) Some imagined that they might not check out the signals with every suggestion but after a period of programming, and in this case, the signal contributed to their overall mental model of copilot's capability and made evaluation easier when they got a similar suggestion in the future: ``it helps with the next piece of code when I use Copilot'' (DP-P5) These signals also made it explicit that there were other users who had faced the same suggestions, giving users a sense of community. For example, DP-P5 stated that they would ``definitely have more trust'' in Copilot with the presence of the signals, as the suggestion has been verified by other users:   

\begin{quote}
    ``Without the feature, I feel like I'm the only one in the dark and Copilot is hovering over behind me. Then with this I see community. I see, hey, there has been 578 other people or at least saw the snippet. I'm like, okay, so these guys used it, then let me just take advantage of it as well.'' (DP-P5) 
\end{quote}

At the same time, participants were concerned that paying attention to the signals could break their programming flow. In DP-P11's word, when looking at the signals, they would ``stop the programming workflow and enter debugging mode.'' Especially in these particular stimuli, users have to click and view the signals in a pop-up window, which was considered disruptive. The participants wished for a representation of the signals that could be digested in a glance. For example, DP-P2 suggested ``a little comment of the code suggestion on the right side saying how other people are responding to this or similar code suggestion.'' They also imagined a hierarchical presentation of information where the user can take a glance at first and dive in to see more if interested. 

In the following section, we elaborate on the user experience and needs with specific concepts in this design. 

\subsubsection{Community usage statistics}
Participants found community usage statistics as a useful, quantified metric that helps them decide the quality of the suggestion: ``it's data-driven... that gives a lot of confidence to users on how good this recommendation is'' (DP-P8). However, the participants were divided on the basis of how they would act on the statistics in actual usage. Some, like DP-P7, believed that ``a higher acceptance percentage'' is ``probably a good sign that I can use this safely.'' Some viewed the number more useful when the numbers were extreme: ``if I see like 90 percent rejected, it gives me some signal,'' but not so much when they were on the middle ground (DP-P1). Others, however, were more skeptical and did not find the numbers directly actionable: ``whether other people have accepted the code, it doesn't have anything to do with my code...[if] it has a bug in my case, it's like 100 percent fail for me.'' (DP-P2)

These divided opinions indicate that users need more scaffolds to interpret any community signals presented in statistics. Besides the numbers, participants would like to see how other users' tasks were relevant to theirs so that they could know how to translate community insights into their use cases: ``the thing that I look for most is how close is their problem to mine.'' (DP-P2) Furthermore, they wished for additional rationales and context behind user action with copilot, as acceptance or rejection along might not show the full picture of user intention. For instance, DP-P4 explained that they tended to experiment with Copilot in their coding process, where ``it was always meant to be rejected because I was just experimenting with it.'' (DP-P4) Therefore, participants would like to see, for example, ``a list of the most common reasons about why the user decided to modify the code or just reject it, or even if the user accepted the code without any change.'' (DP-P3) Information like this can help enhance the user's understanding of the numbers and inform their decisions.

\subsubsection{User identity}
Participants generally appreciated seeing the information of other users to help them evaluate the suggestion. Especially for new programmers, senior developers action on the suggestion could influence their decision. At the same time, some participants also felt that they should not follow others based solely on their organizational titles as presented on this specific design. Additional details on users' experience and expertise directly relevant to the specific suggestions would help with evaluation. For example, feedback from users who ``worked on a similar problem before'' (DP-P2) would be appreciated, as well as their level of experience in the task or programming language. For example, P7 imagined a scenario where they wanted to evaluate the test units suggested by Copilot: ``If I see senior test engineer and they are accepting that. I'm going to probably get good confidence. That could be based on what type of function I'm writing.'' (DP-P7) Additional information on identities can also help users understand where other users' were coming from and the rationales of their actions, further informing users to make effective judgement: ``I try to put in their shoes and see what they were thinking when they liked or didn't like that suggestion.'' (DP-P7)

At the same time, participants wished for a mechanism to protect their identity when contributing data to evaluation signals. For example, DP-P4 demanded an explanation of what data about them was collected and how the data was used: ``they already know my job title, what other information are they collecting about my work? Are they also collecting other information such as code quality and performance?'' (DP-P4) Users needed to ensure that their data is under their control, their privacy is protected, and sensitive data, such as their performance, is not restored.

\subsubsection{Community voting}
Participants appreciated that a user voting mechanism would provide them with an additional dimension that suggested the quality of the suggestion. Compared user actions, the number of likes and shares indicated users' explicit ``registration of immediate satisfaction'' (DP-P10) with the suggested code. Interestingly, a voting mechanism itself could also enhance users' trust with the AI system, as it explicitly indicates that users' ``feedback gets registered'' and their ``voice are heard by the AI'' (DP-P11). For this reason, participants expected to see outcomes of their voting, for example, more similar suggestion to customize their interaction with the AI.

\subsection{Finding 2: User experience and needs with experience sharing spaces}
In general, participants appreciated the dedicated channels for sharing experience with Copilot and engaging in community-curated experience. Compared to the evaluation signals, the participants agreed that they could gain a more in-depth understanding of how others interact with the AI: ``[evaluation signals] showed us statistics. This is one showing us the data samples.'' (DP-P1) These dedicated channels help users determine their trust in AI by providing ``more data points to help to build your own understanding of this Copilot.'' (DP-P5) They would use the community as long as it was ``grassroot'' (DP-P2) and reflected the authentic experience of Copilot users.

At the same time, with this particular design involving video recordings and sharings, the participant hoped that there was a mechanism to make it clear that any sharing and recording is safe and under user consent and control.
For example, DP-P5 demanded the system to `` make it crystal clear that the videos are posted with the owner's consent...and some information about we won't record sensitive information.'' (DP-P5) Furthermore, DP-P6 suggested that the system should provide a feature to warn participants to not share any confidential in their recordings to ensure security and privacy: 

\begin{quote}
``...should put some comment there like when sharing your code, you should verify that you are not going to share in what AWS token or personal information there.'' (DP-P6)
\end{quote}

In the following section, we elaborate how users would use the space within IDE and the external platform differently and their different needs. 

\subsubsection{Experience sharing space within IDE}
When programming inside the IDE, participants tended to leverage community-curated experience when they needed to find strategies for their immediate use case with AI. Some imagined that they would likely check out other users' experience when they were surprised by the suggestions. For participants such as DP-P10, this could include both ``really good or really bizarre suggestions.'' They ``want to see if others have the same experience'' (DP-P10) and a prevalence of sharing could also function as additional evaluation signals to help them decide whether to trust the suggested code. Others view examples shared by others as guidance on how they should interact with the suggestion. For example, DP-P11 shared that they ``will go the community while skeptical about larger piece of suggestions'' to see how others edit the suggestion and how it can translate to their use cases (DP-P11).

On the other hand, since they were in a programming flow, participants felt viewing examples shared by others would be distracted by others' experience. Especially with the current design, the video snippets require    
``maybe too much clicking and time wasting, redirecting my attention too many times'' (DP-P1). Users also have little knowledge about what is in a video and whether it is worth their time:``I don't know what I'm clicking into'' (DP-P2). To address this issue, participants proposed alternatives that could allow users to quickly get the key points from the community-curated experience, including an informative title about the gist of the video snippets (DP-P5), screenshots coupled with a summary of the AI suggestion (DP-P1), snippet of code a few lines before and after the suggestion (DP-P11), or a description of the interaction with a link to the code repository (DP-P11).

\subsubsection{External experience sharing space}
The external experience sharing space can help users build overall trust with AI in aspects such as expectations, strategies, understanding, and awareness of broader impact. 
In contrast to how they would use the within-IDE space, participants imagined themselves using the external space primarily for discovery of general strategies and tips, when they were in their free time instead of in the midst of a coding session. For this reason, they believed that they would most engage with the space when they first learned about the tools: 

\begin{quote}
`` if you see lots of videos that show you all these amazing things, then your expectation will be higher and you saw these tricks, now you want to go try them. Or if you see badly videos, you'll be less likely to use it. It's going to tint how you maybe use it just from this discovery.'' (DP-P1)
\end{quote}

Given their needs in discovery and learning, participants would like to see the experience shared in a form that is more enriched than a recording of a coding session. For example, DP-P8 wished for mechanisms that support adding ``certain concepts'' to the video and make it ``catchy.'' Since they would have time and bandwidth to engage in a video, participants would like the videos to be longer and contain the users' reflection on their experience, as well as a detailed walk-through of the goal, context, and setup of their project: ``people post for a reason, and they should be scaffolded to communicate what problems they are thinking about.'' For these reasons, participants appreciate the discussion feature underneath the videos as they can make sense of the experience with other community members.

Similarly as what we learned from the Study 1 interviews, participants expressed the need to reproduce some of the experience. However, the forking feature in these specific stimuli might not be effective as it could be ``too much investment to fork an entire project just to replicate the interaction'' (DP-P11) and ``the feeling of working on someone else's code, it just feels weird'' (DP-P2). DP-P1 explained the needs as a way to ``extract the resources from the video and try it out...I just want to learn how to use it and take their prompts and paste it onto what I'm doing and seeing if that works.''

\section{Discussion}
In this paper, we present a two-phase study that investigated how online communities shape developers' trust in AI code generation tools. With Study 1 (interviews), we contribute empirical understanding of how community-curated experiences and community-generated evaluation signals help developers build appropriate trust with AI models and what developers seek in this information. To further translate these insights into concrete design recommendations, we contribute with Study 2 a design probe study in which we gathered user experience and needs in a user community integrated into the Copilot experience. Both studies combined demonstrate the importance and design space of introducing community signals to AI code generation systems to facilitate trust. 

\begin{figure*}[t]
  \centering
  \includegraphics[width=0.8\linewidth]{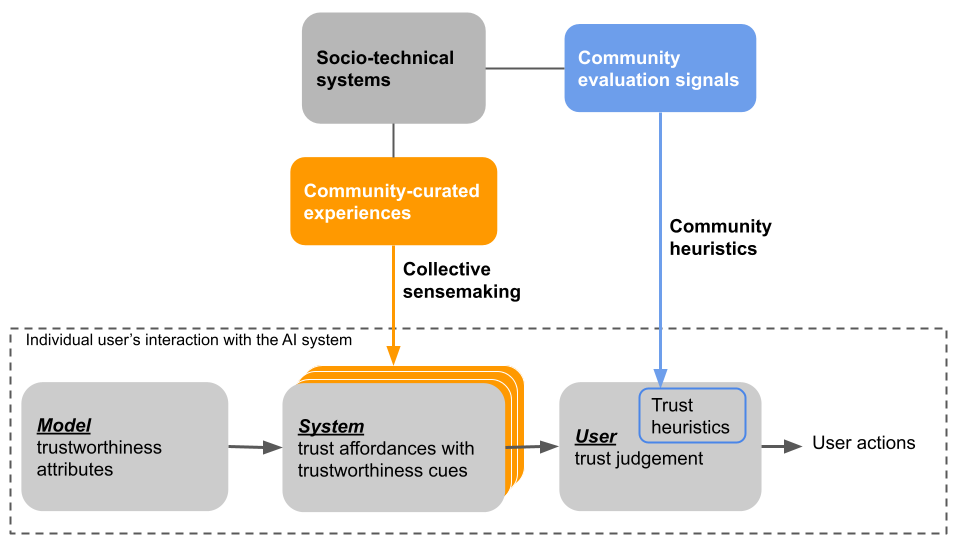}
  \caption{Our study extends the \emph{MATCH model} by surfacing two pathways that sociotechnical systems can help developers build responsible trust with AI tools.}
  \Description{A text box on top representing Social-technical systems is connected to the \emph{MATCH model} flowchart at the bottom via two pathways. The first pathway is connect to the second box in \emph{MATCH model} flowchart, which represents system, trust affordances with trustworthiness cues box, via an arrow labelled as community-curated experiences.The second pathway is connect to the third box in \emph{MATCH model} flowchart, which trust heuristics, via an arrow labelled as community evaluation signals. }
  \label{fig:synthesis}
\end{figure*}

\subsection{Two pathways to building appropriate user trust in AI code generation tools}
\label{sec:discussion_theory}
Echoing previous literature arguing that trust is situated in social technical systems \cite{harawaySituatedKnowledgesScience1988, zhangShiftingTrustExamining2022}, our findings elaborated on the important role of online communities in helping developers build responsible and appropriate trust with AI tools. As a response to the call for social and organizational mechanisms beyond the tool itself to support trust building \cite{liaoDesigningResponsibleTrust2022}, we surfaced two pathways that online communities can offer to help developers build appropriate trust in AI: 1) the pathway of \textit{collective sensemaking} and 2) the pathway of \textit{community heuristics}. Figure \ref{fig:synthesis} shows how these two pathways extend the \emph{MATCH model} by \citet{liaoDesigningResponsibleTrust2022} that we discussed in §\ref{sec:background_trust}. 

The first pathway, \textbf{collective sensemaking}, describes the process by which developers learn from others' experience with the AI model and improve their understanding of the trust affordances of the system. As suggested in previous literature \cite{sarkarWhatItProgram2022}, due to the highly versatile, context-dependent nature of code generation AI, individual developers likely only get exposed to a subset of trust affordances and tend to understand the AI's capacities from their limited perspectives. For example, developers may not be aware of cases where AI can fail, granting too much trust to AI and resulting in mistakes, similar to what we see in the literature on overreliance on and misuse of AI systems~\cite{passiOverrelianceAILiterature, dzindoletRoleTrustAutomation2003, leeTrustAutomationDesigning2004}; or developers may deem AI completely useless only based on their own few negative experiences. With community-curated experiences, developers are exposed to diverse examples of how AI can work or fail in different use cases, which complements their individual understanding of the trust affordances. As a result, developers set
reasonable expectations, learn strategies on when to trust AI, develop empirical understanding of how AI generates suggestions, and develop awareness of broader implications of AI-generated code. All of this can help them make informed decisions about when and how much to trust AI in their programming workflow, helping them build responsible trust \cite{jacoviFormalizingTrustArtificial2021}.

The other pathway, \textbf{community heuristics}, refers to the process in which users leverage a variety of community evaluation signals as heuristics to make trust judgments. Since individual developers rely on their own knowledge, expertise, and intuitions as heuristics, when unfamiliar with the programming project or language, they might not be able to effectively assess the suggested code \cite{vaithilingamExpectationVsExperience2022, sarkarWhatItProgram2022}. Echoing a range of literature on technical Q\&A forums \cite{tausczik2014collaborative, Slag2015OneDayFO, ford2016paradise}, our findings suggest that communities provide crowd-sourced evaluation signals such as user voting, coupled with supplementing discussion context and credibility signals on user identity. These signals can serve as heuristics in addition to the developers' own expertise, scaffolding them to effectively evaluate an AI suggestion and make decisions on whether to trust it. 

\subsection{Design considerations for integrating user community into the AI Code generation experience}
\label{sec:discussion_design}
We present the design space that we constructed based on the findings of our Study 1 interviews that were iterated and enriched according to the feedback of the participants in Study 2. 
Our study highlights the importance of a user community to help developers build appropriate trust with AI code generation tools. 
Even beyond AI for programming support, as more and more commercial applications powered by generative AI model enter people's life, users often have the need to learn from others' interaction with the AI and explore different use cases. In fact, dedicated user communities for a variety of generative AI tools are emerging. For example, users of DALL·E 2 model\footnote{\url{https://openai.com/dall-e-2/}} gather on a platform named DALL·E 2 Gallery\footnote{\url{https://dalle2.gallery/}} to share prompts and the resulting image; some platforms, such as PromptBase\footnote{\url{https://promptbase.com/}}, support users to exchange and sell prompts; \citet{gamage2022deepfakes} investigated a subreddit dedicated to users to make sense of use cases and implications of ``Deepfake'' video synthesis applications\footnote{\url{https://www.theguardian.com/technology/2020/jan/13/what-are-deepfakes-and-how-can-you-spot-them}} powered by generative AI. 
In response to the emerging needs of the user community for generative AI applications, we summarize a series of important design considerations. While the specific recommendations are situated in the context of AI code generation tools, we believe our advice can also be useful for introducing user communities to generative AI applications more broadly.  

\subsubsection{Scaffold users to share specific, authentic experience with AI.}
The premise of a successful user community is the critical mass of user-generated high-quality content \cite{kraut2012building}. Our study suggests that members of a user community dedicated to AI code generation tools favor content about specific experiences with the AI, which can include a detailed description of user interaction with the AI, ideally situated in realistic programming tasks, with sufficient information on project contexts and dependencies. Future platforms should scaffold users to effectively document and share this critical information with the community. Our design probe study suggests that providing auto-recorded video can assist users in sharing their authentic experiences, but has a few important limitations. Future systems can explore other mediums, such as features that help users efficiently create snapshots and informative text-based summarizations of their interaction with the AI. Future platforms can also support users to perform and share metacognitive thinking and ``reflect-in-action'' \cite{schon2017reflective} when interacting with AI suggestions, allowing audiences to further understand their experience.

\subsubsection{Integrate community into users' workflow.}
Getting input from the community when generating content with AI can offer user many benefits, but checking community content in the middle of a task that demands a high mental load such as programming, can be disruptive. Future systems can innovate to reduce such friction. One direction is to innovate on the way community content is presented. For example, information could be presented in a hierarchical manner---users may first take a glance at some sort of indicators, perhaps in the form of icons or numerical scales, that can quickly tell them what is going on in the community about a particular AI suggestion. Users can have the option to dive into the details or otherwise can move on if they want to stay in the task flow. Another possibility is to allow users to customize the timing and frequency of any community signals. Users may not want to see community signals every time they receive a suggestion. Instead, they may only want to see the signals when struggling to evaluate a particular AI suggestion. Future systems could also include mechanisms to detect and predict user intent with AI suggestions and present community signals when needed.

\subsubsection{Assist users to effectively utilize community content.}
Even given great community content and seamless integration into workflow, users may still need support to effectively translate community information into actionable insights. With our specific design probes, users found it particularly difficult to interpret the community usage statistics. Future systems should enable users to understand how community insight can be applied to their specific use cases. Specifically, with community evaluation signals, systems could provide users with concrete examples and context beyond aggregated statistics, such as the common edits community members have made the suggestions and their rationales. Another potential direction could be to signal or visualize the relevance between community content and the user's specific use case, giving users more information on how they should use it.  Furthermore, systems should support users to replicate others' interaction with the AI. This could be features to help users easily extract a prompt from a video post, or perhaps a sandbox mechanism to allow users to play with the prompt in the context of their own project. 

\subsubsection{Assure users about privacy and confidentiality.}
Finally, one of the biggest concerns that users have about the community is about privacy and protection of confidentiality. Especially in high-stake domains such as software engineering, recording and sharing interaction with AI and project context can result in concerns in project confidentiality, such as leaking source code. Although the ``share outside your organization'' switch as part of our design probes is an attempt to address these concerns, users would like additional assurance. Future systems should be aware of these needs and provide mechanisms to ensure users that any formats of recording and sharing are completely under their control. Systems could also give users extra warnings when they share experiences with the public, allow them to preview who may see their content, and perhaps also provide automatic checking on any sensitive and confidential information. Users should also be able to provide informed consent~\cite{imYesAffirmativeConsent2021} on any data about their interaction with AI that is collected, know how it is used, and stop sharing data at any point. 

\subsection{Limitation and future work}

We admit that our work is not without limitations. First, this research was focused on AI-powered code generation tools such as Copilot, Tabnine, Kite, and Amazon Code Whisperer and we intentionally used a variety of AI code generation tools and online communities. However, we cannot claim that the findings will apply to all possible AI applications and to all domains in which AI is applied. We are still confident that community will play a significant role in trust building and also in the success of AI in general, since several communities have already emerged around applications of generative AI in other domains such as creating artwork. Exploring the role of community in building trust for other applications and domains is one direction for future work.

Second, this research is based on 17 interviews with software developers and 11 design probe sessions and could be considered a formative study. Although we have discovered a rich set of insights from our qualitative work, future work should focus on more quantitative analysis of trust and communities with a larger study population. The study was conducted at a time when Copilot was just released and people started adopting Copilot. Therefore, future work should investigate the longitudinal impact of community on trust building.

Third, the demographics of our study sample bias towards a certain gender group. Although we tried our best, we were not able to recruit a pool of participants balanced in gender. Of the 17 participants in Study 1, only two were women, while the rest were all men and no gender-nonconforming participants were recruited. Therefore, it is possible that our insights were biased towards this population and that we might not be able to identify specific experience and needs of others. We want to point out that, in an effort to address the issue of lack of racial diversity in general HCI studies \cite{linxen2021weird}, we were able to recruit a balanced mix of participants in terms of race. Future research can extend our work and test our findings in a broader population of developers. 

Fourth, this research was focused on \emph{current} software developers, and we intentionally included participants with a range of years of programming experience. However, one of the successes of generative AI models has been to broaden who can perform certain tasks. For example, models such as Dall-E and Midjourney allow anyone to be an artist and create digital art. We expect that AI will have similar success in software development and enable more people to make software. Investigating the role of communities and trust for this \emph{next generation} of software developers is another promising direction for future work.

\section{Conclusion}
From our studies, we have been able to confirm that ``trust is a prerequisite''~\cite{eu2019building}
when users are expected to engage with AI systems. Through our work, we have been able to highlight that online communities provide two additional pathways to building appropriate user trust that have not been captured in existing trust models. The pathway of collective sensemaking enhances and complements the user's understanding of trust affordances by sharing others' community-curated experiences with the AI system. The pathway of community heuristics provides users with community evaluation signals to help them make trust judgments. We discussed a series of important design considerations that we believe are relevant for a broad set of generative AI applications that want to leverage the power of community resources.

\bibliographystyle{ACM-Reference-Format}
\bibliography{reference}

\newpage
\appendix

\section{Participant Information}
\begin{table}[ht]
\centering
\caption{The participants of Study 1. The Column \emph{Exp} indicates the years of programming experience. The \emph{Job Title} was self reported by participants.}
\label{tab:study1-participant}
\resizebox{1\textwidth}{!}{%
\small\begin{tabular}{l|p{3cm}p{3.5cm}lp{1.9cm}llp{2.3cm}l}
\textbf{ID} & \textbf{Tool}                                                                                     & \textbf{Community}                                                                                                                       & \textbf{Gender} & \textbf{Race}             & \textbf{Age} & \textbf{Education}  & \textbf{Job Title} & \textbf{Exp} \\ \hline
P1          & GitHub Copilot, Tabnine, Kite, CodeWhisperer, aiXcoder & GitHub (Issues or Community); Discord or Slack; Stack Overflow                                & Man             & Black or African American & 35-44        & Bachelor degree     & IT personnel                     & 16                           \\ 
P2          & GitHub Copilot                                                                                    & Reddit; GitHub (Issues or Community); Discord or Slack                                         & Man             & White                     & 18-24        & Bachelor degree     & Software engineer                & 14                           \\
P3          & None                                                                                              & Reddit                                                                                                                                  & Man             & White                     & 18-24        & High school diploma & Student                          & 5                            \\
P4          & GitHub Copilot                                                                                    & Twitter                                                                                                                                 & Man             & Asian                     & 25-34        & Bachelor degree     & Masters student                  & 8                            \\
P5          & GitHub Copilot, Tabnine                                                                           & GitHub (Issues or Community); Reddit                                                                                                     & Man             & White                     & 25-34        & Bachelor degree     & Software developer/ Analyst      & 5                            \\
P6          & GitHub Copilot, Tabnine, Kite, aiXcoder                                                           & GitHub (Issues or Community); Reddit; Stack Overflow                                                                                      & Man             & Black or African American & 35-44        & Bachelor degree     & Developer                        & 10                           \\
P7          & Tabnine, Kite, CodeWhisperer, aiXcoder                                                            & GitHub (Issues or Community); Reddit; Stack Exchange; Hugging Face Space; Twitter; Stack Overflow & Man             & Black or African American & 25-34        & Bachelor degree     & Developer                        & 10                           \\
P8          & GitHub Copilot                                                                                    & Discord or Slack                                                                                                                        & Man             & Asian                     & 25-34        & Master degree       & Assistant professor              & 10                           \\
P9          & GitHub Copilot                                                                                    & Discord or Slack; Twitter                                                                                                                & Man             & Asian                     & 35-44        & Master degree       & Machine learning\newline engineer        & 10                           \\
P10         & GitHub Copilot                                                                                    & Stack Exchange; Stack Overflow; GitHub (Issues or Community)                                   & Woman           & Asian                     & 25-34        & Master degree       & Student                          & 5                            \\
P11         & GitHub Copilot, Tabnine, CodeWhisperer                                                             & GitHub (Issues or Community); Twitter; Reddit; Stack Overflow                                   & Man             & Black or African American & 25-34        & Bachelor degree     & IT personnel \newline (software)          & 8                            \\
P12         & None                                                                                              & Twitter; Stack Overflow                                                                                                                  & Man             & White                     & 25-34        & Master degree       & PhD student                      & 10                           \\
P13         & GitHub Copilot, Kite                                                                              & Discord or Slack; Hugging Face Space; Reddit; Youtube                                                                                      & Man             & Mix                       & 18-24        & Bachelor degree     & Software engineer                & 3                            \\
P14         & GitHub Copilot, Kite, CodeWhisperer, aiXcoder                                                     & Stack Exchange; Twitter; GitHub (Issues or Community); Reddit; Stack Overflow; Discord or Slack   & Man             & White                     & 25-34        & Bachelor degree     & IT personel                      & 12                           \\
P15         & None                                                                                              & Stack Overflow; Discord or Slack; GitHub (Issues or Community); Youtube                         & Man             & Asian                     & 35-44        & PhD degree          & Software engineer                & 14                           \\
P16         & GitHub Copilot                                                                                    & Twitter; Reddit; Discord or Slack                                                                                                         & Woman           & Asian                     & 25-34        & PhD degree          & ML researcher                    & 14                           \\
P17         & GitHub Copilot                                                                                    & Stack Overflow; GitHub (Issues or Community); Twitter;                                                                                    & Man             & Asian                     & 18-24        & Bachelor degree     & PhD student                      & 5                           \\
\hline
\end{tabular}%
}
\end{table}

\begin{table}[ht]
\centering
\caption{The participants of Study 2. The Column \emph{Exp} indicates the years of programming experience.}
\label{tab:study2-participant}
\resizebox{0.8\textwidth}{!}{%
\small
\begin{tabular}{l|llllll}
\textbf{ID} & \textbf{Copilot}                           & \textbf{Gender} & \textbf{Race}             & \textbf{Age} & \textbf{Education} & \textbf{Exp} \\ \hline
DP-P1       & I use the tool regularly                   & Man             & White                     & 25-34        & Bachelor degree    & 7                            \\
DP-P2       & I use the tool regularly                   & Man             & Asian                     & 18-24        & Master degree      & 6                            \\
DP-P3       & I use the tool regularly                   & Man             & Latino                    & 18-24        & Bachelor degree    & 7                            \\
DP-P4       & I've tried the tool but no longer using it & Woman           & Asian                     & 25-34        & Master degree      & 8                            \\
DP-P5       & I recently started using the tool          & Man             & Asian                     & 25-34        & Bachelor degree    & 7                            \\
DP-P6       & I use the tool regularly                   & Man             & Asian                     & 25-34        & Bachelor degree    & 5                            \\
DP-P7       & I use the tool regularly                   & Man             & White                     & 35-44        & Bachelor degree    & 20                           \\
DP-P8       & I recently started using the tool          & Man             & Asian                     & 18-24        & Bachelor degree    & 7                            \\
DP-P9       & I've tried the tool but no longer using it & Man             & Asian                     & 25-34        & Master degree      & 8                            \\
DP-P10      & I use the tool regularly                   & Man             & Black or African American & 25-34        & Bachelor degree    & 6                            \\
DP-P11      & I recently started using the tool          & Man             & Asian                     & 18-24        & Bachelor degree    & 5                           \\
\hline 
\end{tabular}%
}
\end{table}

\end{document}